\documentclass[12pt,a4paper,oneside]{book}
\usepackage{amsmath}
\usepackage{amssymb}
\usepackage{amsthm}
\usepackage{amsfonts}
\usepackage{graphicx}
\usepackage[all]{xy}
\usepackage{verbatim}
\usepackage[left=2cm,right=2cm,top=3cm,bottom=2.5cm]{geometry}
\usepackage{caption}
\usepackage{subcaption}
\usepackage{float}
\usepackage{stackrel}
\usepackage{setspace}
\usepackage{hyperref}
\usepackage[utf8]{inputenc}
\usepackage[english]{babel}\usepackage{amsmath}
\usepackage[numbers]{natbib}
\usepackage[utf8]{inputenc}
\usepackage[english]{babel}
\usepackage{eucal}
\usepackage{float}
\usepackage{color}

\begin{document}
	
	\pagenumbering{roman}

	\pagestyle{headings}

		\begin{center}
			
			\vspace{0.5in}
			{\bf{Modelling of Bad Biomass Invasion of a Food Chain Ecosystem with Optimal Control.}\\[5mm] Haruna Issaka\textsuperscript{1}, Oluwole Daniel Makinde\textsuperscript{2}, David Mwangi Theuri.\textsuperscript{3}}\\1. Department of Applied Mathematics, Pan Africa University, Institute of Basic Science, Technology and Innovation, Kenya.\\2. Faculty of Military Science, Stellenbosch University, South Africa.\\\underline{3. Faculty of Mathematics, Jomo Kenyatta University of Agriculture, Kenya.}
		\end{center}

	\setcounter{page}{1}
	\pagenumbering{arabic}
	\subsection*{Abstract}
	In this paper we provide a model to describe the dynamics of the species of the ecosystem after it has been raided by a bad competing specie. The competing specie invades the native plants for nutrition, carbon dioxide and space. This affects the population of the native species of the ecosystem.	
	\\
	 The effect of the bad biomass on the ecosystem is examined by considering it's equilibrium points as well as its stability. An optimal control system is developed by using Pontryagin's maximum principle to construct a Hamiltonian function which minimizes the spread of the bad biomass. Numerical simulations were conducted to analyze  the results.   
	It was established that the intrinsic growth rate $r$ of the good biomass was responsible for the sustenance and continuous survival of the ecosystem. If growth rate of the good biomass is increased, the other species showed positive growth. In addition to the growth rate of the good biomass, death rates of both fish and birds also affected the state of the ecosystem. Simulation results also revealed that the invasive bad species biomass introduction affected the growth of the good biomass. This was as a result of the competition for nutrition, carbon dioxide, space between the good and the bad biomasses. After implementing the control units, simulation results showed an improvement in the growth of the good biomass, fish and birds populations whereas it showed a decline in the bad biomass growth.        
	\\
	
	
	\underline{\textit{keywords:} Ecological Modeling, Wetland Invasion, Invading Specie.}

	\subsection*{1. Introduction}
	The ecosystem is one of the major source of salt production in the country providing source of employment and livelihood for majority of the inhabitants around the wetland.
 There are seasonal in-flows of sea water during high tides from the ocean and from rivers such as the Aka river from the north. This inflows helps to maintain water levels to support plants and fish growth \cite{affam2011economic}, \cite{ntiamoa1991seasonal}, \cite{ntiamoa1998water}. Population growth, human activities such as farming, cutting of wood for fuel and climate change has contributed to the reduction of volume of water retention in the lagoon. 
 	In addition, there is also the case of invasion by competing plant species which has displaced a substantial volume of native plants. An invasive species in this regard is a plant specie that can cause an ecological harm in a new environment or ecosystem \cite{masoodi2013predicting}. They are capable of causing total extinction of native plants and animals in the ecosystem. The invasive species, Spartina Alterniflora, is also taking up which was covered by the mangroves which further enhances the depletion of the ecosystem \cite{zheng2016modeling} \cite{issaka2019dynamics}. This invasive species is not consumed by the birds and spreads at a faster rate compared to the growth of the mangroves due to it being monocotyledonous and shallow rooted \cite{issaka2019dynamics}. 	
	Extensive study of wetlands in Ghana done by \cite{affam2011economic} \cite{ntiamoa1998water} has shown that both physical and natural activities have threatened the survival of wetlands in Ghana  and therefore efforts to effectively and efficiently maintain the wetland must be implemented to avoid a complete extinction of the wetland. Both traditional and scientific efforts have been employed to control the depletion of wetlands in Ghana\cite{ntiamoa1998water}. 	
	A wetland consisting of a good biomass, bad biomass and bird population was considered by \cite{upadhyay2014modeling} with Keoladeo national park of India as a case study. In conclusion, they observed that parameter values had a role in determining the dynamics of the wetland. Rai \cite{rai2008modeling} suggested that to ensure good health of the keoladeo national park of India, a constant removal of the bad biomass should be encouraged by allowing harvesting by surrounding communities.   	
	The invasive specie spread by first invading as a non-harmful plant, gradually spread and compete with the native plants and eventually displacing the native plants completely\cite{ge2015plant},\cite{ali1986keoladeo}\cite{upadhyay2014modeling}. When life of good biomass is affected negatively, so does the fish and bird population \cite{issaka2019dynamics}, \cite{sakai2001population}. A reduction in concentration of dissolved oxygen reduces fish and animal population as well as revenue through tourism, depletion of the wetland, drying up of surrounding water bodies amongst the list of associated problems. \cite{ali1986keoladeo},\cite{patra2017modeling},\cite{shukla1998modelling} observed that if the invaded species is removed by any means and the area left to fallow, the species grows again after some time. They therefore suggested a continuous effort in eliminating the bad biomass. 
	\\
	In this paper, we study the interaction between species of the ecosystem using the Beddington-denAgelis functional response and the effect of the invasion on the native plants and by extension to the bird population. Equilibrium points are established and subsequently stability of these equilibrium points if any, examined both locally and globally.		
	\subsection*{2. The model setup}
	We consider a wetland comprising of Prey density (good biomass) $G(t)$, fish population $F$ and birds population (predator) $P(t)$. The good biomass grows by the logistics equation \cite{volterra1928variations} whiles the fish and bird population only grow by interacting with the good biomass. The predator feeds on the the prey by the Beddington-deAgeles functional response \cite{shulin2013dynamics}, \cite{beddington1975mutual}.
	
	We set up the model for the system by the differential equations below:
	\begin{align}
	\label{inter4a1}
	\dfrac{dG}{dt} & = rG\left(1-\dfrac{G}{K}\right) - \dfrac{aGF}{1+c_1G+dF+eP}-\dfrac{bGP}{1+c_2G+dF+eP}-\varepsilon GB\\[2mm]
	\label{inter5a1}
	\dfrac{dB}{dt} & =r_1B\left(1-\dfrac{B}{K_1}\right)-\gamma BF-\sigma BP\\[2mm]
	\label{inter6a1}
	\dfrac{dF}{dt} & = -\alpha F + \dfrac{maGF}{1+c_1G+dF+eP}-\dfrac{fFP}{1+cG+d_1F+eP} -\phi BF\\[2mm]
	\label{inter7a1}
	\dfrac{dP}{dt} & = -\beta P +  \dfrac{nfPF}{1+cG+d_1F+eP}+\dfrac{sbGP}{1+c_2G+dF+eP}-\rho BP
	\end{align}
	
	\noindent where $G(t)\geq0$, $B(t)\geq0$, $F(t)\geq0$ and $P(t)\geq0$ for all $t\geq0$.
	$B(t)$ is the cumulative density of the bad biomass, $r_1$ is the growth rate of bad biomass $\gamma$ and $\sigma$ are the rate of interference between bad biomass and fish, and bad biomass and birds, $\epsilon$  is the rate of interference between bad and good biomass, $K_1$ is the carrying capacity of the bad biomass, $\phi$ and $\rho$ are the death rate of fish and bird population respectively as they interact with bad biomass. All other symbols have their meanings as defined earlier.
	 
	 \subsection*{3. Existence of equilibrium}
	 We obtain the following equilibrium points $E_0(0,0,0,0)$, $E_1(G^*,0,0,0)$, $E_2(G^*,0,F^*,P^*)$,\\ $E_3(G^*,0,F^*,0)$, $E_4(G^*,0,0,P^*)$, $E_5(G^*,B^*,0,0)$, $E_6(G^*,B^*,F^*,0)$, $E_7(G^*,B^*,0,P^*)$,\\ $E_8(G^*,B^*,F^*,P^*)$.
	 \\
	 $1.$ The equilibrium point of $E(G,B,0,0)$ is $(K(1-\dfrac{\varepsilon}{r}K_1), K_1)$  for $K_1<\dfrac{r}{\varepsilon}$ 
	 \\
	 $2.$ The equilibrium point of $E_6$ is obtained by decomposing the system into $f(g,B)$ and $g(G,B)$ and graphically solving for $G^*, B^*$. The values of $G^*$ and $B^*$ is the point of intersection of $f$ and $g$. Knowing the $G^*, B^*$, the values of $F^*$ and $P^*$ can be calculated.\\ Similar analogy is used to obtained the rest of the equilibrium points.
	 
	 \section*{4. Stability of Equilibrium points}
	  \textbf{Local Stability}\\ The Jacobian matrix is defined by\\
	 $J=\begin{pmatrix}
	 	j_{11}	& j_{12} & j_{13} & j_{14} \\[10pt] 
	 	j_{21}	& j_{22} & j_{23} & j_{24} \\[10pt] 
	 	j_{31}	& j_{32} & j_{33} & j_{34} \\[10pt]
	 	j_{41}  & j_{42} & j_{43} & j_{44}
	 	\end{pmatrix}$
	where\\[4mm]
	$j_{11}=r(1-\dfrac{2G}{K})-\dfrac{(1+dF+eP)(aF+bP)}{(1+cG+dF+eP)^2}-\varepsilon B$, \quad
	$j_{12}=-\varepsilon G$,
	\\[3mm]
	$j_{13}=\dfrac{G(bdP-a(1+cG+eP))}{(1+cG+dF+eP)^2}$, 
	\quad
	$j_{14}=\dfrac{G(aeF-b(1+cG+dF))}{(1+cG+dF+eP)^2}$,
	\quad
	$j_{21}=0$,
	\\[3mm]
	$j_{22}=r_1(1-\dfrac{2B}{K_1})-\gamma F-\sigma P$, 
	\quad
	$j_{23}=-\gamma B$,
	\quad
	$j_{24}=-\sigma B$,
	\quad
	$j_{31}=\dfrac{F(cfP+am(1+dF+eP))}{(1+cG+dF+eP)^2}$,
	\\[3mm]
	 $j_{32}=-\phi F$,
	\quad
	$j_{33}=\dfrac{(1+cG+eP)(amG-fP)}{(1+cG+dF+eP)^2}-\alpha -\phi B$,
	\quad
	$j_{34}=-\dfrac{F(aemG-f(1+cG+dF))}{(1+cG+dF+eP)^2}$,
	\\[3mm]
	$j_{41}=\dfrac{P(sb(1+dF+eP)-cnfF)}{(1+cG+dF+eP)^2}$,
	\quad
	$j_{42}=-\rho P$,
	\quad	  
	$j_{43}=\dfrac{P(nf(1+cG+dF)-bdsG)}{(1+cG+dF+eP)^2}$, 
	\\[3mm]
	$j_{44}=\dfrac{(1+cG+dF)(nfF+bsG)}{(1+cG+dF+eP)^2}-\beta -\rho B$
	\\[3mm]
	$\textbf{Theorem}$ $1.$ The system $E_5$ is locally asymptotically stable if $r<\varepsilon$.
	\\
	\textbf{Proof} For the system $E_5$ to be stable, the eigenvalue, $\lambda<0$.
	\\
	Simplifying and solving for $\lambda$ from the above matrix, we get:
	\\
	$((r(1-\dfrac{2G}{K})-\varepsilon B)-\lambda)(-bG-\lambda)=0$.
	$\lambda_1=r(1-\dfrac{2G}{K})-\varepsilon B$ and $\lambda_2=-bG$. Hence $E_5$ is stable.
	\\
	$2.$ The characteristic equation associated with $E_6$ is
	
\noindent	$\lambda^3-(j_{11}+j_{22}+j_{33})\lambda^2+(j_{11}j_{22}+j_{11}j_{33}+j_{22}j_{33}-j_{23}j_{32}-j_{13}j_{31})\lambda+(j_{13}j_{31}j_{22}-j_{12}j_{23}j_{31}-j_{11}j_{22}j_{33})=0$ or equivalently, $\lambda^3+a_1\lambda^2+a_2\lambda+a_3=0$, where
	$a_1=-(j_{11}+j_{22}+j_{33})$, $a_2=(j_{11}j_{22}+j_{11}j_{33}+j_{22}j_{33}-j_{23}j_{32}-j_{13}j_{31})$ and $a_3=(j_{13}j_{31}j_{22}-j_{12}j_{23}j_{31}-j_{11}j_{22}j_{33})$.\\ Applying the Routh-Hurwitz criteria \cite{pal1979stable},\cite{s1945simplification}, $E_6$ is stable if $a_1>0$, $a_3>0$ and $a_1a_2>a_3$, which implies $i.$ $r<\varepsilon$ and $G^*<\dfrac{K}{2}$
	$ii.$ $r_1<\gamma$ and $B^*<\dfrac{K_1}{2}$
	$iii.$ $f>a$
	\\
	$3.$ The characteristic equation associated with $E_7$ is 
$\lambda^3-(j_{11}+j_{22}+j_{44})\lambda^2+(j_{11}j_{22}+j_{11}j_{44}+j_{24}j_{42}-j_{41}j_{44})\lambda+(j_{11}j_{24}j_{42}+j_{22}j_{41}j_{44}-j_{11}j_{22}j_{44}-j_{22}j_{24}j_{41})=0$. or equivalently 
\\
$\lambda ^3+a_1\lambda^2+a_2\lambda+a_3=0$ where 
\\
$a_1=-(j_{11}+j_{22}+j_{44})$, $a_2=(j_{11}j_{22}+j_{11}j_{44}+j_{24}j_{42}-j_{41}j_{44})$ and $a_3=(j_{11}j_{24}j_{42}+j_{22}j_{41}j_{44}-j_{11}j_{22}j_{44}-j_{22}j_{24}j_{41})$. By the Routh-Hurwitz criteria, $E_7$ is stable if $a_1>0$, $a_3>0$ and $a_1a_2>a_3$ which can easily be verified.
\\
$4.$ The associated characteristic equation is \quad
$\lambda^4-(j_{11}+j_{22}+j_{33}+j_{44}+j_{34}j_{43})\lambda^3+(j_{11}j_{22}+j_{11}j_{33}+j_{11}j_{44}+j_{22}j_{33}+j_{22}j_{44}+j_{33}j_{44})\lambda^2-(j_{11}j_{22}j_{33}+j_{11}j_{22}j_{44}+j_{11}j_{33}j_{44}-j_{22}j_{34}j_{43}-j_{11}j_{34}j_{43})\lambda+(j_{11}j_{22}j_{33}j_{44}-j_{11}j_{22}j_{34}j_{43}-j_{22}j_{33}j_{44})=0$, \quad
which is represented as \quad $a_0\lambda^4 +a_1\lambda^3 + a_2\lambda^2 + a_3\lambda + a_4 =0$.
\\
The characteristic equation has a finite number of terms of order $4$ and coefficient of the first term is non zero i.e. $a_0> 0$. To show from the characteristic equation whether or not the system is stable using the Routh array, we need to show that there is no sign change in moving from one term to another. If there exist a sign change, then the number of times the sign changes determines the number of positive eigen values present. This indicates an unstable equilibrium point.   
We form the Routh array as in the table below\\
\begin{center}
	\begin{tabular}{ |p{3cm}|p{4cm} p{2cm} p{2cm}| }
		\hline
		\multicolumn{4}{|c|}{Routh Array} \\
		\hline
		$\lambda$ & Coefficients \\
		\hline
		$\lambda^4$ & $a_0$ & $a_2$ & $a_4$ \\
		$\lambda^3$ & $a_1$   & $a_3$ & $0$ \\
		$\lambda^2$ & $\dfrac{a_1a_2-a_0a_3}{a_1}$ & $\dfrac{a_1a_4}{a_1}$ & $0$ \\
		$\lambda$  & $\dfrac{a_1a_2a_3-a_0a_3^2-a_1^2a_4}{a_1a_2-a_0a_3}$ & $0$ & $0$ \\
		$\lambda^0$ & $a_4$ & $0$ & $0$ \\
		\hline
	\end{tabular}
\end{center}   
 \noindent
The necessary and sufficient condition for the eigen values to be negative is that all the $a_i's$ must be positive i.e. $(a_i>0)$ and all entries in the first column be positive. 

In this case for stability, we ensure that 

\begin{enumerate}
	\item $\dfrac{a_1a_2-a_0a_3}{a_1}>0. \quad\quad i.e. \quad a_1a_2>a_0a_3.$\\
	\item $\dfrac{a_1a_2a_3-a_0a_3^2-a_1^2a_4}{a_1a_2-a_0a_3}>0. \quad\quad i.e. \quad a_1a_2a_3>a_0a_3^2+a_1^2a_4.$
\end{enumerate}
\noindent
If any of these conditions is not satisfied, the system becomes unstable.

\section*{5. Optimal Control System.} We formulate the control to the system of the state equations (system of Differential Equations) using the Pontryagin's principle. Since the aim of the research is to minimize the spread of the invasive species and the cost of the control while maximizing the growth of the native population, we consider the maximization principle of Pontryagin \cite{pontryagin2018mathematical}.

\section*{6. Control Interventions}
These are measures employed to control the spread of the bad biomass after they invade the wetland ecosystem. Various methods can be applied to control the spread and we shall consider a few below.

\begin{enumerate}
	\item \textbf{Physical Control:} This method comprises the use of manual means to control the invaded species from spreading. These methods include hand picking, selective weeding, burning of affected areas etc. This method is time consuming and does not take long for the weeds to start growing again.
	\item \textbf{Chemical Control:} The chemical control of weeds requires the use of techniques in the application of herbicides to weeds or the soil to control the growth or germination of the weed species and the growth of the bad biomass. It is effective in certain cases and has a wide scope depending on the cost, efficiency and the availability of the chemical. The herbicides either helps in killing the weeds or inhibit their growth. Selective herbicides are the most effective for the control of weeds.
	\item \textbf{Biological Control:}  This involves the use of natural species such as plants or animals to control the germination of bad weeds. It is a new trend of weed control. The bio-agents feeds on only the selected weeds and not on the entire vegetation. This is an effective and efficient method provided the right type of predators  are introduced under starvation conditions and will not feed on the good biomass.
	\item \textbf{Quarantine or Isolation:} Prevention is the most effective method of dealing with invasive species. Once an invasive has entered an area and become established, eradication is far more expensive and it is likely that greater resources will be required to control its further spread and reduce its impact.The first step in preventing invasion is to prevent the entry of the invasive.
	
	Quarantine is the protection of unaffected wetlands from coming into contact with the invasive species. This process can be achieved through creating barriers, trenches or the construction of road as a barrier around the unaffected species.
\end{enumerate}

\noindent We present an optimal control equation using Pontryagins maximum principle by incorporating into equations the system of differential equations, the control variable $u_i(t)$ where $i=1,2,3,4$ and $0\leq u_i(t)\leq 1$. 
\\
Consider the system of differential equations modelling an ecosystem under invasion:
\begin{align}
\label{inter8}
\dfrac{dG}{dt} & = rG\left(1-\dfrac{G}{K}\right) - \dfrac{aGF}{1+c_1G+dF+eP}-\dfrac{bGP}{1+c_2G+dF+eP}-\varepsilon GB\\[2mm]
\label{inter9}
\dfrac{dB}{dt} & =r_1B\left(1-\dfrac{B}{K_1}\right)-\gamma BF-\sigma BP\\[2mm]
\label{inter10}
\dfrac{dF}{dt} & = -\alpha F + \dfrac{maGF}{1+c_1G+dF+eP}-\dfrac{fFP}{1+cG+d_1F+eP} -\Phi BF\\[2mm]
\label{inter11}
\dfrac{dP}{dt} & = -\beta P +  \dfrac{nfPF}{1+cG+d_1F+eP}+\dfrac{sbGP}{1+c_2G+dF+eP}-\rho BP
\end{align}

\noindent where $G(t)\geq0$, $B(t)\geq0$, $F(t)\geq0$ and $P(t)\geq0$ for all $t\geq0$.
\\
The behavior of the underlying dynamics of the system is described by the state variables $G, B, F, P$ and we assume that there is a way to monitor the state variable by acting on it with a suitable control function. The control enters the system and affects the dynamics of the state variables. Our aim is to adjust the control to minimize the spread of the bad biomass and hence improve the growth of the good biomass.
\\
Now we adjust the control functions $u_i(t)$, for all $0\leq u_i(t)\leq 1$ to the desired state. $u_1$, represents the rate of removal of the bad biomass as a proportion to the density of the effort and bad biomass by means of curative (biological) control measures. $u_2$ represents effort in treating affected parts of the ecosystem by physical or chemical means, $u_3$ represents effort in quarantining the unaffected parts of ecosystem, $u_4(t)$ is the effort applied in improving the growth of the good biomass.
If the control function $u_1(t)$ is close to $1$, then the removal rate of the bad biomass is high but with a high cost of implementation.
\\
From the model defined above, the Objective function to be minimized is

\begin{equation}
\min\limits_{0\leq u_i\leq1}J(u_i)=\int_{0}^{t}(z_0B(t)+\sum_{i=1}^{4}\dfrac{z_i}{2}u_i^2(t))dt,
\end{equation}

\noindent subject to:

\begin{align}
\label{inter4}
\dfrac{dG}{dt} & = rG\left(1-\dfrac{(1-u_3)G}{K}\right) - \dfrac{aGF}{1+c_1G+dF+eP}-\dfrac{bGP}{1+c_2G+dF+eP}- (1-u_2)\varepsilon GB\\[2mm]
\label{inter}
\dfrac{dB}{dt} & =r_1B\left(1-\dfrac{(1-u_1)B}{K_1}\right)-\gamma BF-\sigma BP\\[2mm]
\label{inter6}
\dfrac{dF}{dt} & = -\alpha F + \dfrac{maGF}{1+c_1G+dF+eP}-\dfrac{fFP}{1+cG+d_1F+eP} -(1-u_4)\Phi BF\\[2mm]
\label{inter7}
\dfrac{dP}{dt} & = -\beta P +  \dfrac{nfPF}{1+cG+d_1F+eP}+\dfrac{sbGP}{1+c_2G+dF+eP}-(1-u_4)\rho BP
\end{align}

\noindent where $G(t)\geq0$, $B(t)\geq0$, $F(t)\geq0$, $P(t)\geq$ for all $t\geq0$, $z_0>0$, $z_1>0$, $z_2>0$	  \\
and $z_0$ represents the cost of the spread of the bad biomass, $\frac{z_1}{2}u_1^2(t)$ is the cost of effort to reduce the spread of the invasive species at any time t, $\frac{z_2}{2}u_2^2(t)$ is the cost of effort in treating affected areas of the good biomass, $\frac{z_3}{2}u_3^2(t)$ represents the effort in quarantining the unaffected areas of the good biomass from invasion and $\frac{z_4}{2}u_4^2(t)$ is the cost of effort in protecting the fish and bird species from interacting with the bad biomass.
\\
The aim therefore is to find an optimal control tuple  $(u_1^*,u_2^*,u_3^*,u_4^*)$ \\such that\\ $$J(u_1^*,u_2^*,u_3^*,u_4^*)=\min\limits_{0\leq u_i\leq1}J(u_1,u_2,u_3,u_4)$$. 
\\
To do this, we apply the pontryagin's principle to the above Objective function and its associated state equations (constraints) by transforming it into a Hamiltonian function \cite{lenhart2007optimal}, \cite{runge2002importance}.
The Hamiltonian function is generated as below with the assumption that $c=c_1=c_2$, $d=d_1$.

$H(t,G,B,F,P,\lambda) =[z_0B(t)+\dfrac{z_1}{2}u_1^2(t)+\dfrac{z_2}{2}u_2^2 +\dfrac{z_3}{2}u_3^2+\dfrac{z_4}{2}u_4^2 ] +\\
\lambda_1\left[rG\left(1-\dfrac{(1-u_4)G}{K}\right) -\dfrac{aGF}{1+cG+dF+eP} - \dfrac{bGP}{1+cG+dF+eP}-\varepsilon (1-u_2)GB\right]\\
+\lambda_2\left[r_1B\left(1-\dfrac{(1-u_1)B}{K_1}\right)-\gamma BF-\sigma BP\right]\\
+\lambda_3\left[-\alpha F + \dfrac{maGF}{1+cG+dF+eP}-\dfrac{fFP}{1+cG+dF+eP} -\phi (1-u_3) BF\right]\\
+\lambda_4\left[-\beta P +  \dfrac{nfPF}{1+cG+dF+eP}+\dfrac{sbGP}{1+cG+dF+eP}-\rho (1-u_3) BP\right]$
\\
with associated adjoint equations as\\
$\dfrac{d\lambda_1}{dt}=-\dfrac{\partial H}{\partial G},\quad \dfrac{d\lambda_2}{dt}=-\dfrac{\partial H}{\partial B},\quad\dfrac{d\lambda_3}{dt}=-\dfrac{\partial H}{\partial F},\quad\dfrac{d\lambda_4}{dt}=-\dfrac{\partial H}{\partial P}.$
\\
computed as

\begin{align*}
\begin{tiny}
\begin{cases}
\dfrac{d\lambda_1}{dt}=\lambda_1\left(\dfrac{2(1-u_4)G}{K}-r+\varepsilon (1-u_2)B\right)-\dfrac{[1+df+ep][aF(\lambda_3 m-\lambda_3)+bP(\lambda_4 s-\lambda_1)]}{(1+cG+dF+eP)^2}\\[3mm] 
\dfrac{d\lambda_2}{dt}=-z_0-\lambda_2r_1+\lambda_1 \varepsilon (1-u_2)G+\dfrac{2\lambda_2 r_1(1-u_1)B}{K_1}+(\lambda_2r_1+\lambda_3\phi (1-u_3))F+(\lambda_2\sigma +\lambda_4\rho(1-u_3))P\\[3mm]
\dfrac{d\lambda_3}{dt}=\lambda_3\alpha+(\lambda_3 \phi (1-u_3)-\lambda_2\gamma)B-\dfrac{1}{(1+c_2G+dF+eP)^2}[ac_2(\lambda_3m-\lambda_1)G^2+a(\lambda_3m-\lambda_1)G+\\(ae(\lambda_3m-\lambda_1)+cf(\lambda_3+\lambda_4n)+bd(\lambda_1-\lambda_4s)]GP+f(\lambda_4n-\lambda_3)P+ef(\lambda_4n+\lambda_3)P^2.\\[3mm]
\dfrac{d\lambda_4}{dt}=\lambda_2\sigma B+\lambda_4\beta +\lambda_4\rho (1-u_3)B-\dfrac{1}{(1+cG+dF+eP)^2}[c(bs\lambda_4-\lambda_1a)G^2+(\lambda_4 bs-\lambda_1 a)G+\\(\lambda_1 a(e-d)+cf(\lambda_4 n-\lambda_3)+(\lambda_4bds-\lambda_3aem))GF+f(\lambda_4 n-\lambda_3)F+df(\lambda_4 n-\lambda_3)F^2].
\end{cases}
\end{tiny}
\end{align*}

\noindent
and the optimality condition is given by
$\dfrac{\partial H}{\partial u_i}=0$:

\begin{align*}
\dfrac{\partial H}{\partial u}=
\begin{cases}
\dfrac{\partial H}{\partial u_1}=z_1u_1+\dfrac{\lambda_2 r_1B^2}{K_1}\\[2mm]
\dfrac{\partial H}{\partial u_2}=z_2u_2+\lambda_1 \epsilon GB\\[2mm]
\dfrac{\partial H}{\partial u_3}=z_3u_3+\lambda_3 \phi BF + \lambda_4 \rho BP\\[2mm]
\dfrac{\partial H}{\partial u_4}=z_4u_4+\dfrac{\lambda_1 rG^2}{K}
\end{cases}
\end{align*}

\begin{align*}
u^*=
\begin{cases}
u_1^*(t)=-\dfrac{\lambda_2 r_1B^2}{z_1K_1}\\[2mm]
u_2^*(t)=-\dfrac{\lambda_1 \epsilon GB}{z_2}\\[2mm]
u_3^*(t)=-\dfrac{\lambda_3 \phi BF-\lambda_4 \rho BP}{z_3}\\[2mm]
u_4^*(t)=-\dfrac{\lambda_1 rG^2}{z_4K}
\end{cases}
\end{align*}
with transversality condition $\lambda_i(T_f)=0$.\\	  
Characterizing the cost functions $(u_1^*,u_2^*,u_3^*,u_4^*)$, we have
\begin{align*}
u^*(t)=
\begin{cases}
u_1^*(t)=max\lbrace 0, min(1, -\dfrac{\lambda_2 r_1B^2}{z_1K_1} \rbrace\\
u_2^*(t)=max\lbrace 0, min(1, -\dfrac{\lambda_1 \epsilon GB}{z_2} \rbrace\\
u_3^*(t)=max\lbrace 0, min(1, -\dfrac{\lambda_3 \phi BF-\lambda_4 \rho BP}{z_3} \rbrace\\
u_4^*(t)=max\lbrace 0, min(1, -\dfrac{\lambda_1 rG^2}{z_4K} \rbrace\\
\end{cases}
\end{align*}

\begin{align*}
=\begin{cases}
u_1^*(t)=max\lbrace 0, min(1, \tau_1)\rbrace\\
u_2^*(t)=max\lbrace 0, min(1, \tau_2)\rbrace\\
u_3^*(t)=max\lbrace 0, min(1, \tau_3)\rbrace\\
u_4^*(t)=max\lbrace 0, min(1, \tau_4)\rbrace
\end{cases}
\end{align*}

where 
\\
$\tau_1=-\dfrac{\lambda_2 r_1B^2}{z_1K_1}$, $\tau_2=-\dfrac{\lambda_1 \epsilon GB}{z_2}$, $\tau_3=-\dfrac{\lambda_3 \phi BF-\lambda_4 \rho BP}{z_3}$ and 
\\
$\tau_4=-\dfrac{\lambda_1 rG^2}{z_4K}$.\\[3mm]
\\
The optimal system is formed from the state system and the adjoint system by incorporating the control set together with the initial and transversality conditions.

\begin{align*}
\begin{tiny}
\begin{cases}
\dfrac{dG}{dt}  = rG\left(1-\dfrac{(1-u_3)G}{K}\right) - \dfrac{aGF}{1+c_1G+dF+eP}-\dfrac{bGP}{1+c_2G+dF+eP}- (1-u_2)\varepsilon GB\\[2mm]
\dfrac{dB}{dt}  =r_1B\left(1-\dfrac{(1-u_1)B}{K_1}\right)-\gamma BF-\sigma BP\\[2mm]
\dfrac{dF}{dt}  = -\alpha F + \dfrac{maGF}{1+c_1G+dF+eP}-\dfrac{fFP}{1+cG+d_1F+eP} -(1-u_4)\Phi BF\\[3mm]
\dfrac{dP}{dt}  = -\beta P +  \dfrac{nfPF}{1+cG+d_1F+eP}+\dfrac{sbGP}{1+c_2G+dF+eP}-(1-u_4)\rho BP\\[3mm]
\dfrac{d\lambda_1}{dt}=\lambda_1\left(\dfrac{2(1-u_4)G}{K}-r+\varepsilon (1-u_2)B\right)-\dfrac{[1+df+ep][aF(\lambda_3 m-\lambda_3)+bP(\lambda_4 s-\lambda_1)]}{(1+cG+dF+eP)^2}\\[3mm] 
\dfrac{d\lambda_2}{dt}=-z_0-\lambda_2r_1+\lambda_1 \varepsilon (1-u_2)G+\dfrac{2\lambda_2 r_1(1-u_1)B}{K_1}+(\lambda_2r_1+\lambda_3\phi (1-u_3))F+(\lambda_2\sigma +\lambda_4\rho(1-u_3))P\\[3mm]
\dfrac{d\lambda_3}{dt}=\lambda_3\alpha+(\lambda_3 \phi (1-u_3)-\lambda_2\gamma)B-\dfrac{1}{(1+c_2G+dF+eP)^2}[ac_2(\lambda_3m-\lambda_1)G^2+a(\lambda_3m-\lambda_1)G+\\(ae(\lambda_3m-\lambda_1)+cf(\lambda_3+\lambda_4n)+bd(\lambda_1-\lambda_4s)]GP+f(\lambda_4n-\lambda_3)P+ef(\lambda_4n+\lambda_3)P^2.\\[3mm]
\dfrac{d\lambda_4}{dt}=\lambda_2\sigma B+\lambda_4\beta +\lambda_4\rho (1-u_3)B-\dfrac{1}{(1+cG+dF+eP)^2}[c(bs\lambda_4-\lambda_1a)G^2+(\lambda_4 bs-\lambda_1 a)G+\\(\lambda_1 a(e-d)+cf(\lambda_4 n-\lambda_3)+(\lambda_4bds-\lambda_3aem))GF+f(\lambda_4 n-\lambda_3)F+df(\lambda_4 n-\lambda_3)F^2].
\end{cases}
\end{tiny}
\end{align*}

\section*{7. Numerical Simulation}
In this section, the dynamical behaviour of the proposed system model $4$ is discussed using MATLAB 2018a and MAPLE 2018. Data used for the simulations and its analysis was obtained from the Center for Scientific and Industrial Research (CSIR), Ghana and shown below:\\ $\alpha= 0.05;\quad \beta=0.06; \quad a= 0.2; \quad K= 12; \quad b= 0.8; \quad c= 0.3; \quad c_1= 0.3; \\ \quad c_2= 0.3; \quad e= 0.5; \quad d= 0.4; \quad r= 1.7; \quad f= 0.01; \quad m= 2; \quad n=2; \\ \quad s=0.5; \quad d_1= 0.02; \quad K_1=12; \quad r_1=1; \quad \gamma=0.03; \quad \sigma=0.03; \quad \varepsilon=0.04; \\ \quad \phi=0.045; \quad \rho=0.05;$
\\
From chapter $3$, we were able to show that the positive equilibrium point $E(G,F,P)$ is locally and globally stable under desirable conditions and we conclude that all the species coexisted simultaneously. After the invasion of the ecosystem by the bad biomass, the stability dynamics of the system is affected as can be seen in figure $5.3(a)$. The new state of the system after invasion, shows an initial oscillatory motion before attaining stability with the bad biomass enjoying a steady growth by displacing the good biomass. 
\begin{figure}[H]
	\centering
	\begin{subfigure}[H]{0.5\textwidth}
		\includegraphics[width=\textwidth]{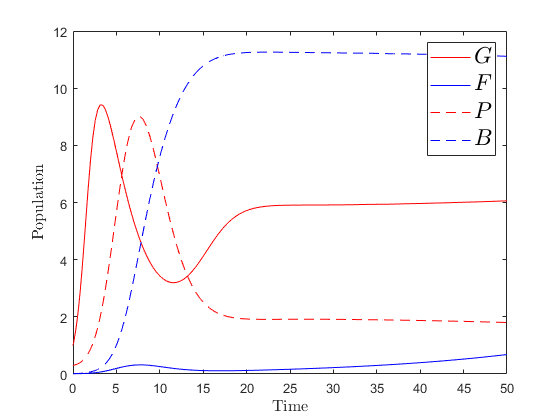}
		\caption{State of ecosystem after invasion.}
	\end{subfigure}
	~ 
	\begin{subfigure}[H]{0.4\textwidth}
		\includegraphics[width=\textwidth]{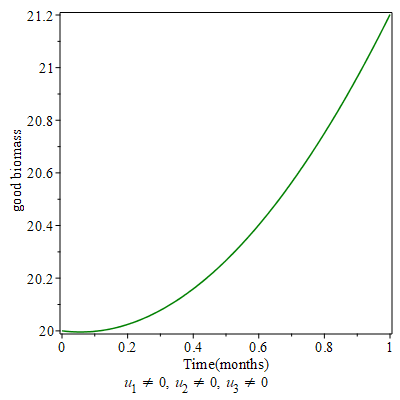}
		\caption{State of ecosystem after control implementation.}
	\end{subfigure}
	\caption{State of ecosystem before and after Optimal Control.}
\end{figure}

\begin{figure}[H]
	\centering
	\begin{subfigure}[H]{0.45\textwidth}
		\includegraphics[width=\textwidth]{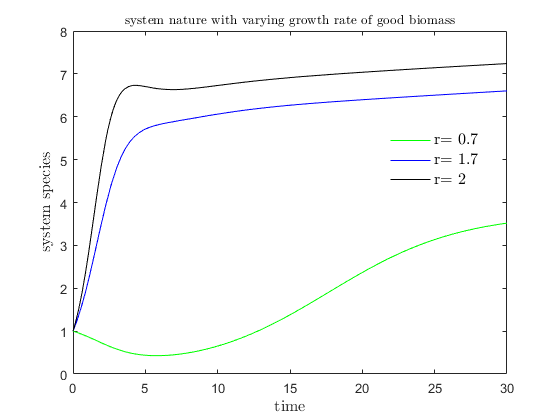}
		\caption{Ecosystem with varying growth rate $r$ of good biomass.}
	\end{subfigure}
	~ 
	\begin{subfigure}[H]{0.45\textwidth}
		\includegraphics[width=\textwidth]{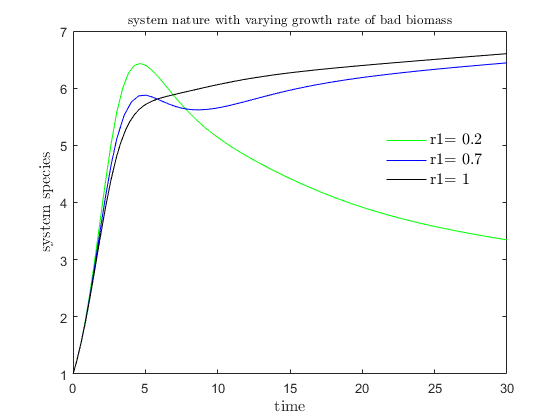}
		\caption{Ecosystem with varying growth rate $r_1$ of bad biomass.}
	\end{subfigure}
	\caption{State of ecosystem before and after Optimal Control.}
\end{figure}
\noindent	  	  
A substantial amount of the density of the good biomass, fish and bird populations is lost a result of interacting with the bad biomass. To be able to avert this, we study the behaviour of the system by varying some parameters and observe the nature of the system. From figure $5.4(a)$ and with the above parameter values, if we increase the growth rate $r$ of the good biomass and keeping all other parameters constant, the system shows an improvement in the growth of the good biomass with a corresponding increase in the fish and birds population.
\begin{figure}[H]
	\centering
	\begin{subfigure}[H]{0.45\textwidth}
		\includegraphics[width=\textwidth]{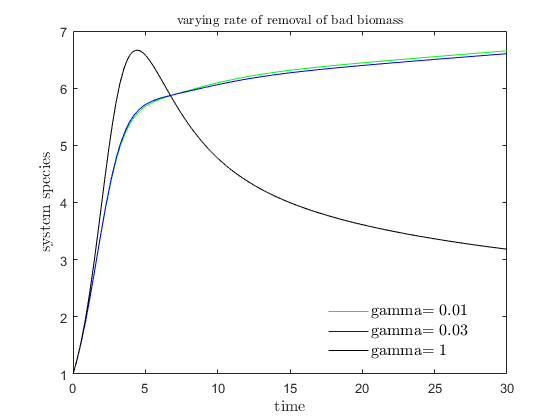}
		\caption{Removal rate $\gamma$ of bad biomass.}
	\end{subfigure}
	~ 
	\begin{subfigure}[H]{0.45\textwidth}
		\includegraphics[width=\textwidth]{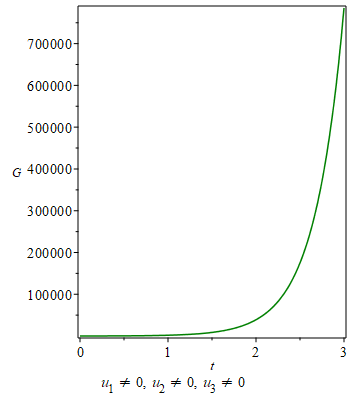}
		\caption{Good biomass performance after optimal control.}
	\end{subfigure}
	\caption{Effect of removing bad biomass on good biomass.}
\end{figure}		  	  
\noindent Figure $5.4(b)$ shows a decrease in the system density with increasing growth rate $r_1$ of bad biomass over time. This implies if there is continuous removal of the bad biomass by effort, (physical, chemical or biological means), will decrease the spread thereby improving growth in good biomass. This is shown in figure $5.5(a)$. We see that as $\gamma$ increases, the total density of the system increases. 
\begin{figure}[H]
	\centering
	\begin{subfigure}[H]{0.45\textwidth}
		\includegraphics[width=\textwidth]{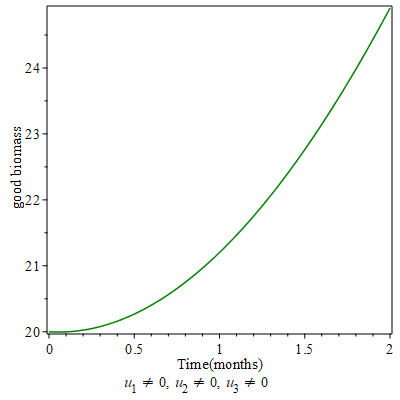}
		\caption{Good biomass with Control.}
	\end{subfigure}
	~ 
	\begin{subfigure}[H]{0.45\textwidth}
		\includegraphics[width=\textwidth]{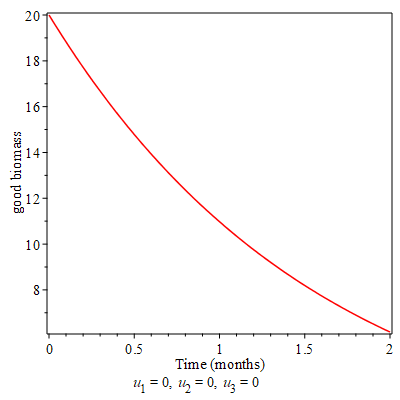}
		\caption{Good biomass without control.}
	\end{subfigure}
	\caption{State of good biomass before and after Optimal Control.}
\end{figure}	  	  
\noindent In addition to the parameters listed above, we set $z_0=22,\quad z_1=3,\quad z_2=5,\quad z_3=2$ and $z_4=3$, $G(0)=0.5$, $F(0)=0.1$ and $P(0)=0.1$. We perform numerical simulations on the optimal control problem of the system equations to observe the effect of the implementations of the various control strategies. We apply the \cite{lenhart2007optimal} approach to solve the Hamiltonian system formed from the Pontryagins maximum principle. The process involves solving the optimal system of the state and the co-state equations using the Runge-Kutta iterative scheme. Considering the initial conditions of the state equations and the final condition of the adjoint equations, we solve the state equation by the forward fourth order Runge-Kutta method and the co-state equations which are final conditions, by the backward fourth order. We adjust the updated control using a combination of the previous values and values from the characterizations of $u_i^*$\quad $\forall i=1,2,3,4.$ The process is repeated until the unknown at the current iteration is sufficiently close to the previous.
\\
In line with this, we proposed the following control strategies to curb the spread of the bad biomass and to restore affected areas to it's original use.
\begin{enumerate}
	\item $u_1$ represents the rate of removal of the bad from affected areas by introduction of herbivorous animals to graze.
	\item $u_2$ is effort in providing treatment to affected parts using chemicals, ploughing etc.
	\item $u_3$ effort in providing quarantine to unaffected areas.
	\item $u_4$ effort applied to restoring affected areas by replanting native plants in affected areas.
\end{enumerate} 
\begin{figure}[H]
	\centering
	\begin{subfigure}[H]{0.45\textwidth}
		\includegraphics[width=\textwidth]{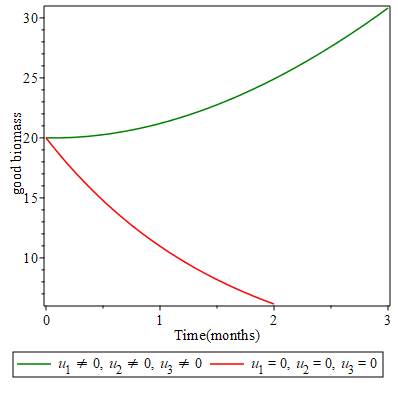}
		\caption{Good biomass before and after control intervention.}
	\end{subfigure}
	~ 
	\begin{subfigure}[H]{0.45\textwidth}
		\includegraphics[width=\textwidth]{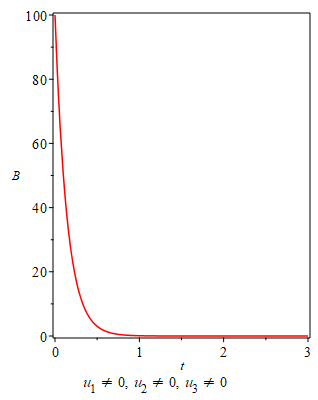}
		\caption{Bad biomass performance after optimal control.}
	\end{subfigure}
	\caption{State of ecosystem before and after Optimal Control.}
\end{figure}	
\textbf{Control with removal of bad biomass} 
\\
Simulation results shows that continuous removal of bad biomass by allowing grazing at affected areas of the ecosystem improves the quality and growth of the good biomass. Figure $5.5(b)$ shows an initial growth which starts slowly but improves and eventually increases growth with time. The growth in the good biomass is exponential showing clearly that if removal of bad biomass is carried out over a long period, the ecosystem will regain back its status.  We then suggest that continuously allowing animals to graze on the affected areas will improve the good health of the ecosystem.
\\
\textbf{Control effort using chemical and ploughing}
\\
From figure $5.6a$, the use of chemical or ploughing increases the growth of the good biomass. This method decreases the rate of competition between the good biomass and the bad biomass thereby increasing the growth of the good biomass as well as fish and bird populations. figure $5.6b$ is a graph showing the state of the good biomass before the implementation of the controls. We see a decline in the growth of the good biomass due to competition from the bad biomass for space, oxygen and nutrition. This is an indication that the so long as the invasive specie continues to exist in the ecosystem, it will eventually displace the good biomass and lead the ecosystem to extinction. In figure $5.7a$, we compare the states of the good biomass before and after the control interventions.
\\
\textbf{Effect of control on bad biomass}
\\
After the control interventions are applied, we monitor the performance of the bad biomass. Simulation results showed a decline in the total biomass of the bad biomass as is shown in figure $5.7b$. The result shows an effectiveness of the control units selected. 
\begin{figure}[H]
	\centering
	\begin{subfigure}[H]{0.45\textwidth}
		\includegraphics[width=\textwidth]{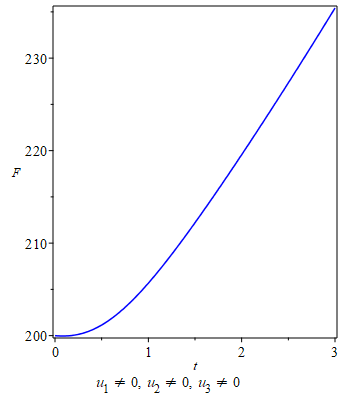}
		\caption{Growth of bird population after control implementation.}
	\end{subfigure}
	~ 
	\begin{subfigure}[H]{0.45\textwidth}
		\includegraphics[width=\textwidth]{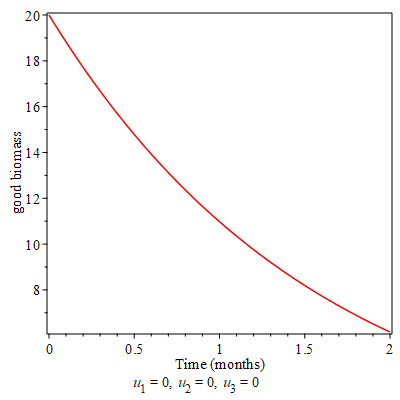}
		\caption{Good biomass performance after optimal control $(u_1\neq0,u_2\neq0,u_3\neq0)$.}
	\end{subfigure}
	\caption{State of ecosystem before and after Optimal Control.}
\end{figure}	
\noindent
\textbf{Fish population after control interventions}
\\
A positive growth in the density of the fish population is observed after the control intervention as can be seen from figure $5.8a$. As indicated earlier, the growth of the fish population is dependent on the amount of the good biomass consumed by the fish. Invariantly, a growth in the good biomass will result in the growth of the fish population.

\section*{8. Discussions}
In this paper, we proposed a model to study the dynamics of ecosystem invaded by a bad biomass competing with and displacing the good biomass. Positive and feasible equilibrium point were established as well as stability of feasible equilibrium points determined. We observed that once the wetland is invaded, there is a shift in parameter values. Some parameter values where affected negatively more than others. For instance the growth rate of the good biomass decreased drastically after the invasion, which had an adverse effect on the reproduction of fish and birds. After examining the effect of the introduction of the bad biomass on the good biomass as well as on the fish and bird population, we modeled a control system using Pontryagin's maximum principle to avert the anormally. The state of the ecosystem improved after the control mechanism was implemented. The growth of the good biomass improved along side the fish and bird populations. We observed that once the removal of the bad biomass is increased, its growth decreases and this enhances the growth of the good biomass with the release of enough dissolved oxygen in the water bodies to improve aquatic life.

	  \renewcommand{\bibname}{References}
	  \bibliographystyle{abbrvnat}
	  \bibliography{references}
	  \addcontentsline{toc}{chapter}{References}
	  \nocite{*}
\end{document}